# Clustering Method for Time-Series Images Using Quantum-Inspired Computing Technology

Tomoki Inoue[1], Koyo Kubota[1], Tsubasa Ikami[2], Yasuhiro Egami[3], Hiroki Nagai[2], Takahiro Kashikawa[4], Koichi Kimura[4], Yu Matsuda[1, *]

1. Department of Modern Mechanical Engineering, Waseda University, 3-4-1 Ookubo, Shinjuku-ku, Tokyo, 169-8555, Japan

2. Institute of Fluid Science, Tohoku University, 2-1-1 Katahira, Aoba-ku, Sendai, Miyagi-prefecture 980-8577, Japan

3. Department of Mechanical Engineering, Aichi Institute of Technology, 1247 Yachigusa, Yakusa-Cho, Toyota, Aichi-prefecture 470-0392, Japan

4. Quantum Application Core Project, Quantum Laboratory, Fujitsu Research, Fujistu Ltd, Kanagawa 211-8588, Japan

* corresponding author: Yu Matsuda

**Email:** y.matsuda@waseda.jp

**Author Contributions:**

Tomoki Inoue: Conceptualization, Data curation, Formal analysis, Investigation, Methodology, Software, Validation, Visualization.

Koyo Kubota: Formal analysis, Investigation, Methodology, Software, Validation, Visualization

Tsubasa Ikami: Data curation, Investigation, Resources, Visualization, Writing -review & editing

Yasuhiro Egami: Data curation, Formal analysis, Investigation, Resources, Writing -review & editing

Hiroki Nagai: Investigation, Resources, Writing -review & editing

Yasuo Naganuma: Methodology, Software, Validation

Koichi Kimura: Methodology, Software, Validation

Yu Matsuda: Conceptualization, Data curation, Formal analysis, Funding acquisition, Investigation, Methodology, Project administration, Resources, Software, Supervision, Validation, Visualization, Writing – original draft, Writing – review & editing

**Competing Interest Statement:** Takashiro Kashikawa and Koichi Kimura are employees of Fujitsu Ltd.

**Classification:** Physical Sciences, Computer Science

**Abstract**

Time-series clustering serves as a powerful data mining technique for time-series data in the absence of prior knowledge about clusters. A large amount of time-series data with large size has been acquired and used in various research fields. Hence, clustering method with low computational cost is required. Given that a quantum-inspired computing technology, such as a simulated annealing machine, surpasses conventional computers in terms of fast and accurately solving combinatorial optimization problems, it holds promise for accomplishing clustering tasks that are challenging to achieve using existing methods. This study proposes a novel time-series clustering method that leverages an annealing machine. The proposed method facilitates an even classification of time-series data into clusters close to each other while maintaining robustness against outliers. Moreover, its applicability extends to time-series images. We compared the proposed method with a standard existing method for clustering an online distributed dataset. In the existing method, the distances between each data are calculated based on the Euclidean distance metric, and the clustering is performed using the k-means++ method. We found that both methods yielded comparable results. Furthermore, the proposed method was applied to a flow measurement image dataset containing noticeable noise with a signal-to-noise ratio of approximately 1. Despite a small signal variation of approximately 2%, the proposed method effectively classified the data without any overlap among the clusters. In contrast, the clustering results by the standard existing method and the conditional image sampling (CIS) method, a specialized technique for flow measurement data, displayed overlapping clusters. Consequently, the proposed method provides better results than the other two methods, demonstrating its potential as a superior clustering method.

**Introduction**

The collection of large-sized datasets has significantly increased with advancements in data storage and data acquisition technology. Time-series data containing one or multiple variables (e.g., image) varying over time is extensively recorded and analyzed in various fields, such as science, engineering, medical science, economics, finance. [1-3] Clustering is a powerful data mining technique for classifying these data into related groups in the absence of sufficient prior knowledge about the groups. [4-6] In particular, when dealing with time-series data, the clustering technique is referred to as time-series clustering. [7-9] Many studies on time-series clustering have been conducted, which are summarized in review papers. [2,7-10] Additionally, some libraries for time-series clustering have been made available on the web [11-15] and are widely used. Following the literature, [7,8] time-series clustering is defined as "the process of unsupervised partitioning a given time-series dataset into clusters, in such a way that homogenous time-series are grouped together based on a certain similarity measure, is called time-series clustering." Three main methods are commonly employed for time-series clustering: raw-data-based/shape-based, feature-based, and model-based approaches. [7,8] These methods differ in their initial calculation procedures. The raw-data-based/shape-based approach directly uses the raw data for clustering, while the feature-based approach transforms the raw data into a low dimensional feature vector. The model-based approach assumes that the time-series data is generated from a stochastic process model, and the parameters of the model are estimated from the data. Since the performance of model-based approach degrades when clusters are close to each other, raw-data-based and feature-based approaches are usually used. [2,8] The subsequent step involves calculating the similarity or distance between two data, feature-vectors, models. Then, the data is grouped into clusters based on the measured similarity or distance using machine learning methods. Clustering algorithms commonly employed for time-series data include partitioning, hierarchical, model-based, and density-based clustering algorithms. [7,8] Among partitioning clustering algorithms, k-means clustering is one of the most widely used algorithms. [5,6,16,17] Its main advantage lies in its low computational cost. However, the method requires a number of clusters to be pre-determined by users. In a hierarchical clustering algorithm, number of clusters does not need to be pre-determined. However, once clusters are split or merged using the divisive or agglomerative method, they cannot be adjusted. Neural network approaches such as self-organizing maps [18] and hidden Markov model [5] are used as model-based clustering approaches. In addition to the above-mentioned disadvantage of the performance degradation for the close clusters, these approaches suffer from high computational cost. Density-based methods, such as density-based spatial clustering of applications with noise (DBSCAN), [19] do not need to pre-determine number of clusters and is robust to outliers.

However, in these methods, an appropriate choice of parameters is difficult, and it is known that they suffer from the curse of dimensionality.

This study proposes a novel time-series clustering method using an annealing machine as an alternative clustering approach. Annealing machines, such as quantum annealing and simulated annealing machines, solve combinatorial optimization problems faster and more accurately than conventional computers. [20-23] Therefore, we expect that our proposed method can achieve clustering tasks that are challenging to achieve with existing methods. One unique characteristic of the proposed method is its ability to evenly classify time-series data into closely related clusters while maintaining robustness against outliers. Additionally, the method was specifically designed to evenly classify periodic time-series images into several phase ranges because of assuming a sufficient number of images for each phase, given the long duration of the time-series data relative to the period. This paper provides a comprehensive explanation of our proposed method. We used the third-generation Fujitsu Digital Annealer (DA3), which is one of the quantum-inspired computing technologies, for the clustering computations. DA3 can solve quadratic unconstrained binary optimization (QUBO) problems, and the clustering problem can be formulated as an Ising model which is equivalently a QUBO problem. [24,25] DA3 provides a solution in large-scale problem space of up to 100 kbits. Subsequently, we applied our proposed method to two time-series datasets: one obtained from "the UEA & UCR time-series classification repository," [26-28] and the other consisted of flow measurement image data capturing the Kármán vortex street obtained in our previous data. [29-31] We specifically chose flow measurement data because it was typically high-dimensional ($\sim 10^6$) and contained a measurement noise. For the clustering process, we employed the raw-data-based and the feature-based approaches. Furthermore, we compared our results with those obtained from standard existing methods, specifically "tslearn" [11] available online, and the conditional image sampling (CIS) method [32,33] (only for flow measurement data).

**Results and Discussion**

### Clustering of online available time-series dataset

We demonstrated the application of the proposed method to classify the "crop" dataset available from the UEA & UCR time-series classification repository. [26-28] The clustering results obtained using the "TimeSeriesKMeans" function in "tslearn" and the proposed methods are shown in Fig.

1. The "crop" dataset contained twenty-four clusters. However, we present the results of two representative clusters. In this dataset, the correct classifications were known and displayed in Fig. 1. Additionally, ensemble-averaged data for each method were calculated. As shown in Fig. 1(a), the proposed method successfully classified the data, while the results obtained by the standard existing method (tslearn) exhibited some unfavorable classifications. We calculated the root mean squared error (RMSE) between the correct data and the results obtained by the proposed method and "tslearn". The RMSEs of the proposed method and the existing method shown in Fig. 1(a) were 0.013 and 0.015, respectively. This further confirmed that the proposed method surpassed the standard existing method. On the other hand, the RMSEs of the proposed and the existing methods shown in Fig. 1(b) were 0.013 and 0.009, respectively. In this condition, the result obtained by the proposed method was inferior to that of the existing method. However, since the variance of the correct data is large as shown in Fig. 1(b), the classification is inherently difficult. Consequently, we can conclude that the proposed method provides results comparable to those of conventional methods.

### Clustering of flow measurement time-series dataset

We applied our method to the flow measurement dataset to demonstrate its effectiveness for noisy data. A typical data of a snapshot is shown in Fig. 2, and the image shows that the data contains noticeable noise with a signal-to-noise ratio of approximately 1. In this study, we classified this time-series data into nine clusters using the proposed method, "tslearn", and the CIS method. The clustering results are shown in Fig. 3, where the data is presented on a two-dimensional scatter plot using multi-dimensional scaling (MDS). In the MDS calculation, the distance between the data $\mathbf{x}_i$ and $\mathbf{x}_j$ is represented as $|\sin(\theta_{i,j}/2)|$, where $|a|$ represents the absolute value of $a$, and $\theta_{i,j}$ corresponds to the angle between data vectors $\mathbf{x}_i$ and $\mathbf{x}_j$. Since the Kármán vortex street dataset used here is a periodical phenomenon with a maximum distance of unity, the data were distributed along a circle with a radius of $1/2$. As illustrated in Fig. 3, the proposed method successfully classified the data without overlaps. The data points were evenly classified into each cluster, and the cluster sizes were similar to each other. The data points out the circle with the radius of $1/2$ were considered outliers, which is a reasonable because these data were considered disturbances deviating from periodic phenomena. However, the outliers were classified into one of the clusters in the standard existing method. This will be inappropriate when calculating ensemble averaging of the data. In the CIS method, only the data on the circle are classified. However, some clusters exhibited overlapping regions and did not form discrete clusters. Density-based methods, such as density-based spatial clustering of applications with noise (DBSCAN), are known as powerful clustering methods. However, the data points on the circle were classified into a single cluster in DBSCAN.

The ensemble-averaged pressure distributions are shown in Fig. 4. The proposed method (Fig. 4a) and the CIS method (Fig. 4c) well extract a periodic vortex generation despite a small pressure variation of approximately 2%. On the other hand, the pressure distribution obtained from the standard method failed to extract the periodic motion accurately. For example, the vortex located at the upper side suddenly disappeared from phase 2 to phase 3, and the vortex at the upper side reversed its flow direction from phase 5 to phase 6 (Fig. 4b). This discrepancy is attributed to the overlapping clusters observed in Fig. 3b. As the pressure decreases when the vortex comes, we compared the minimum pressure at the center of the vortex between the proposed method and the CIS method. The ensemble-averaged pressure values were $p/p_{\text{ref}} = 0.982 \pm 0.001$ and $p/p_{\text{ref}} = 0.984 \pm 0.002$ for the proposed and the CIS methods, respectively, where the error represents the standard deviation and $p_{\text{ref}}$ denotes the atmospheric pressure. The pressure by the CIS method was slightly higher than that of the proposed method, which aligned with the observations in Fig. 4a and c. This difference indicates that the vortex was weakened in the CIS method due to overlapping clusters mentioned earlier, where data from different phases were also included in the ensemble averaging process. These findings provide further evidence that the proposed method is a powerful clustering approach for analyzing periodic phenomena.

**Materials and Methods**

## Proposed method for time-series clustering

We propose a novel clustering method using an annealing machine. We focused on the raw-data-based and the feature-based approaches for time-series data analysis. We considered a clustering problem that a given dataset of $n$ time-series data $\mathbf{X} = \{\mathbf{x}_1, \mathbf{x}_2, \cdots, \mathbf{x}_n\}$, where $\mathbf{x}_i$ is a column vector, is classified to $k$ clusters $\mathcal{C} = \{c_1, c_2, \cdots, c_k\}$. The Hamiltonian for the clustering problem is described as follows: [34,35]

$$\mathcal{H}_1 = \sum_{\mathcal{C}} \sum_{i<j} d_{i,j}\, q_{g,i} q_{g,j} - \lambda_1 \sum_{\mathbf{X}} \left( \sum_{\mathcal{C}} q_{g,j} - 1 \right)^2, \tag{1}$$

where $q_{g,i} = 1$ when $\mathbf{x}_i$ belongs to cluster $c_g$ and $q_{g,i} = 0$ when $\mathbf{x}_i$ does not belong to the cluster $c_g$, that is,

$$q_{g,i} = \begin{cases} 1 : \mathbf{x}_i \in c_g \\ 0 : \mathbf{x}_i \notin c_g \end{cases}. \tag{2}$$

The similarity or inverse of the distance between $\mathbf{x}_i$ and $\mathbf{x}_j$ is denoted as $d_{i,j}$, and $\lambda_1$ is a hyperparameter. The sum $\sum_{i \neq j} d_{i,j} q_{g,i} q_{g,j}$ represents the sum of the similarity or the inverse of the distance between two data points belonging to a cluster. The sum $\sum_{\mathcal{C}}$ represents the sum over all clusters in the first term of Eq. 1. Clustering can be calculated by minimizing $-\mathcal{H}_1$, i.e.,

$$\min -\sum_{\mathcal{C}} \sum_{i<j} d_{i,j} q_{g,i} q_{g,j} + \lambda_1 \sum_{\mathbf{X}} \left( \sum_{\mathcal{C}} q_{g,j} - 1 \right)^2. \tag{3}$$

The second term in Eq. 3 represents a constrained term ensuring each data point belongs to only one cluster. [34,35] The value $\lambda_1$ determines the strictness of this constraint, where smaller value allows for the possibility of some data points being treated as outliers and not assigned to any cluster. This study considered the following minimization problem:

$$\min -\sum_{\mathcal{C}} \sum_{i<j} d_{i,j} q_{g,i} q_{g,j} + \lambda_1 \sum_{\mathbf{X}} \left( \sum_{\mathcal{C}} q_{g,j} - 1 \right)^2 + \lambda_2 \sum_{\mathcal{C}} \left( \sum_{j} q_{g,j} \right)^2, \tag{4}$$

where the new term, the third term in Eq. 4, adjusts the number of data points classified into each cluster. We denote $S_g = \sum_j q_{g,j}$ to simplify the notation, indicating the number of data points belonging to the cluster $c_g$. Then, the third term of Eq. 4 is written as

$$\sum_{\mathcal{C}} \left( \sum_j q_{g,j} \right)^2 = \sum_{\mathcal{C}} S_g^2. \tag{5}$$

The mean number of data points and the variance of data points belonging to each cluster are represented by $\mu$ and $\sigma^2$, respectively. Equation 5 is written as

$$\begin{aligned} \sum_{\mathcal{C}} S_g^2 &= \sum_{\mathcal{C}} \left\{ (S_g - \mu)^2 + 2 S_g \mu - \mu^2 \right\} \\ &= k(\sigma^2 + \mu^2), \end{aligned} \tag{6}$$

where the number of the data points $n$ and the clusters $k$ are fixed, and $\mu = k/n$ is a constant. Then, as the variance decreases, i.e., the third term in Eq. 4 becomes smaller, the data points are evenly classified into each cluster.

### Time-series dataset for demonstration

We applied the proposed clustering method to two time-series datasets. One of the datasets, named "crop," was obtained from the UEA & UCR time-series classification repository. [26-28]

These time-series data were derived from the images taken by the FORMOSAT-2 satellite. The dataset consists of 24 classes corresponding to an agricultural land-cover map, and each data corresponds to its temporal evolution. The time-series length was 46, and the number of dimensions of the data was 1. The data was standardized to have a mean of 0 and a variance of 1. We compared the clustering results obtained by the proposed method and those obtained by "tslearn" [11] In this study, we used the "TimeSeriesKMeans" function in "tslearn." The parameters in the function were set to general settings as follows: the number of clusters was 24, the metric (distance between each data) was Euclidean, the method for initialization was k-means++, and the other parameters were employed default values. This is a standard time-series clustering method. In the proposed method, the Euclidean distance was also used as the metric, and the inverse of the metric was used to minimize the first term in Eq. 4. Since all data points should belong to one of the clusters in this dataset, the parameter $\lambda_1$ was approximately 100 times larger than $\lambda_2$. In this condition, the solution that all data point belonged to one of the clusters (the second term of Eq. 4 was 0) was obtained.

The second dataset used in this study was the flow image data obtained in our previous study [29-31] measuring using the pressure-sensitive paint (PSP) method. [36-38] PSP is a pressure distribution measurement technique based on the oxygen quenching of the phosphorescence emitted from the dyes incorporated into the PSP coating. The measured data was the pressure distribution induced by the Kármán vortex street behind a square cylinder. The data size was $780 \times 780$ spatial grids. The flow velocity was 50 m/s, and the Reynolds number was $1.1 \times 10^5$. The pressure difference was too small to detect the PSP technique due to the small variation in the phosphorescence intensity. Then, the measured pressure contained a noticeable noise, and the noise should be reduced from the data. It is well known that the Kármán vortex is a periodic phenomenon. The data was classified into several phase groups and averaged within these groups to reduce the noise and to extract significant patterns, which is one of the purposes of time-series clustering. The cosine similarity measure was used to assess the similarity between the data because we focused on the phase information of the vortex. Since the PSP data was a time-series image data with two spatial dimensions and one temporal dimension, the pressure distribution data was reshaped into a column vector. Consequently, the time-series PSP data is written as $n$ time-series data $\mathbf{X} = \{\mathbf{x}_1, \mathbf{x}_2, \cdots, \mathbf{x}_n\}$, where $\mathbf{x}_i$ is a vector corresponding to a reshaped pressure distribution. Since the measured PSP data contains a significant noise, the denoised data was used for the calculation of the similarity. Following the literature, [39] the dataset with small noise can be obtained by considering the truncated singular value decomposition (SVD). We considered a data matrix $\mathbf{Y} =$ $[\mathbf{x}_1 \ \mathbf{x}_2 \ \cdots \mathbf{x}_n]$, where the data matrix $\mathbf{Y}$ is obtained by arranging vectors $\mathbf{x}_i$ in time-series order. SVD provides the following representation:

$$\mathbf{Y} = \mathbf{U}\mathbf{\Sigma}\mathbf{V}^\mathsf{T}, \tag{7}$$

where the matrices $\mathbf{U}$ and $\mathbf{V}$ are unitary matrices, and the superscript $\mathsf{T}$ shows the transpose. The matrix $\mathbf{\Sigma}$ is a diagonal matrix of singular values arranged in descending order. It is well known that the data can be approximated by a truncated SVD [40] as follows:

$$\widetilde{\mathbf{Y}} = \widetilde{\mathbf{U}}\widetilde{\mathbf{\Sigma}}\widetilde{\mathbf{V}}^\mathsf{T}, \tag{8}$$

where $\widetilde{\mathbf{\Sigma}}$ is a first $r \times r$ diagonal matrix and $r$ is a truncation rank. The matrices $\widetilde{\mathbf{U}}$ and $\widetilde{\mathbf{V}}$ are reduced matrices corresponding to $\widetilde{\mathbf{\Sigma}}$. Then, we obtained the noise-reduced time-series data matrix of $\widetilde{\mathbf{Y}} = [\widetilde{\mathbf{x}}_1 \ \widetilde{\mathbf{x}}_2 \ \cdots \widetilde{\mathbf{x}}_n]$ or the time-series data of $\widetilde{\mathbf{X}} = \{\widetilde{\mathbf{x}}_1, \widetilde{\mathbf{x}}_2, \cdots, \widetilde{\mathbf{x}}_n\}$. We set $r = 5$, which is a commonly used truncation value. Subsequently, the cosine similarity $\cos\theta_{i,j}$ was calculated as follows:

$$\cos\theta_{i,j} = \frac{\langle \widetilde{\mathbf{x}}_i, \widetilde{\mathbf{x}}_j \rangle}{\|\widetilde{\mathbf{x}}_i\|_2 \|\widetilde{\mathbf{x}}_j\|_2}, \tag{9}$$

where $\langle \widetilde{\mathbf{x}}_i, \widetilde{\mathbf{x}}_j \rangle$ is the inner product of $\widetilde{\mathbf{x}}_i$ and $\widetilde{\mathbf{x}}_j$, and $\|\widetilde{\mathbf{x}}_i\|_2$ is the $\ell 2$ norm of $\widetilde{\mathbf{x}}_i$. This study only considered the pressure distribution behind the square cylinder to reduce computational cost. Substituting $d_{i,j} = \cos\theta_{i,j}$ in Eq. 4, we calculated the clustering using DA3. The parameter $\lambda_1$ was several ten times larger than $\lambda_2$, ensuring that each term in Eq. 4 was of a similar magnitude. In this condition, some data were classified as outliers. The images within the same cluster were ensemble averaged to extract significant patterns. Here, we note that the original image data $\mathbf{X}$ was averaged to extract the patterns, while the truncated dataset of $\widetilde{\mathbf{X}}$ was not used.

Considering that $\cos\theta_{i,j}$ lies within the range of $-1$ to $1$, we introduced the following relation $r_{i,j}$, which range from of $0$ to $1$:

$$\begin{aligned} r_{i,j} &= \frac{\cos\theta_{i,j} + 1}{2} \\ &= \cos^2\frac{\theta_{i,j}}{2} \\ &= 1 - \sin^2\frac{\theta_{i,j}}{2}, \end{aligned} \tag{10}$$

where $1 - r_{i,j} = 0$ when $i = j$ (the same data). Then, we define the distance metric between the data as $|\sin(\theta_{i,j}/2)|$, where $|a|$ represents the absolute value of $a$ and $\theta_{i,j}$ is the angle between data vectors. The maximum value of the distance is unity in this distance metric. This distance metric was used in the MDS calculation.

The time-series data were also classified by the "TimeSeriesKMeans" function in "tslearn" described above. Additionally, we used the CIS method [32,33], which is a specialized methods designed specifically for PSP measurements. In the CIS method, the time-series data were classified into several phase groups based on the pressure data measured by a pressure transducer sensor which is a point sensor with a higher sampling rate than PSP. In other words, the CIS method requires an additional sensor for clustering. This reliance on an extra sensor can be considered one of the limitations of the CIS method.

**Conclusions**

We propose a novel clustering method using an annealing machine. We added the new term that adjusts the number of data classified into each cluster to a QUBO model. In this study, we applied our proposed method to two distinct datasets: one is the "crop" dataset available from the UEA & UCR time-series classification repository and the other is a flow measurement image dataset obtained in our previous study. For the clustering of "crop" dataset, we also employed a standard existing method distributed as "tslearn," in which the distance between each data was calculated based on the Euclidean distance and the clustering was calculated by the k-means++ algorithm. Comparing the results obtained from our proposed method and the existing method, we observed that the variation of the data points obtained by the proposed method was smaller than that by the existing method. In this dataset, the correct clustering result was provided. Then, we calculated the ensemble averaged data, and the root mean squared errors (RMSEs) between the correct data and the ensemble averaged data were compared. Our findings indicate that both methods provide similar result for this dataset.

Next, we applied our clustering method to the flow measurement image dataset which consisted of the time-series pressure distributions induced by the Kármán vortex street. This dataset exhibited periodicity. Another characteristic of this data is that the dataset contains a noticeable noise with a signal-to-noise ratio of approximately 1. For comparison, the dataset was also classified using the standard existing method and the conditional image sampling (CIS) method, which is specifically designed for flow measurement data. The proposed method successfully classified the data without any overlap between the clusters in spite of the small pressure variation of approximately 2%. On the other hand, both the existing and the CIS methods exhibited overlapping of clusters, failing to form discrete clusters. In particular, the overlap between the clusters calculated by the existing method was large; thus, the vortex suddenly disappeared at times and exhibited reverse flow at other times in the ensemble-averaged pressure distribution. It was also found that the vortex was weakened in the ensemble-averaged pressure distribution obtained by the CIS method. These outcomes highlight the superior performance of the proposed method in the clustering periodic phenomena.

**Acknowledgments**

The authors would like to express their gratitude to Dr. Yasuhumi Konishi, Mr. Hiroyuki Okuizumi and Mr. Yuya Yamazaki for their assistance during the wind tunnel testing conducted at the Institute of Fluid Science, Tohoku University. The authors would also like to acknowledge Mr. Takafumi Oyama for the insightful discussions. We also gratefully appreciate Tayca Corporation for providing the titanium dioxide. We would like to thank Editage (www.editage.com) for English language editing. Part of the work was conducted under the Collaborative Research Project J23I041 of the Institute of Fluid Science, Tohoku University.

**Figures and Tables**

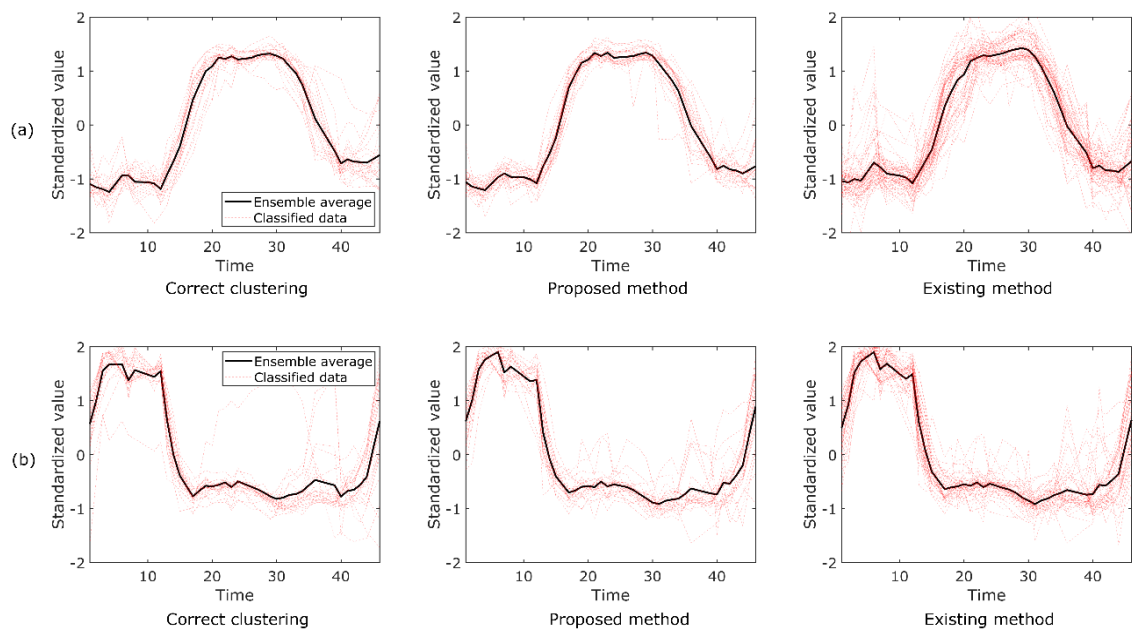

**Figure 1.** Typical clustering results for "crop" dataset from the UEA & UCR time-series classification repository using the proposed method and the existing method. The data labeled as class 1 and class 17 in the repository are shown in (a) and (b), respectively.

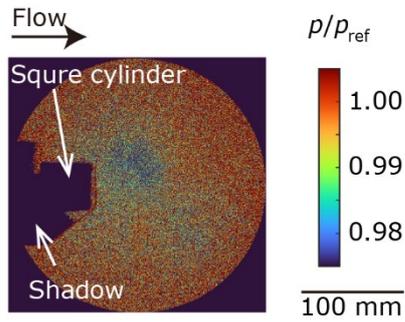

**Figure 2.** Typical pressure distribution, where pressure $p$ is normalized by an atmospheric pressure $p_{\text{ref}}$. Reproduced from Inoue et al. [31].

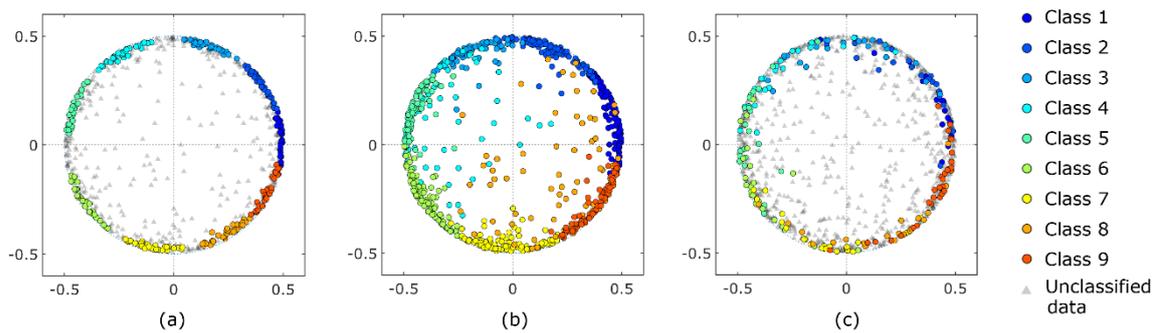

**Figure 3.** Clustering results shown in two-dimensional scatter plot based on MDS. (a) the result by the proposed method, (b) that by the existing standard method (tslearn), (c) that by the CIS method.

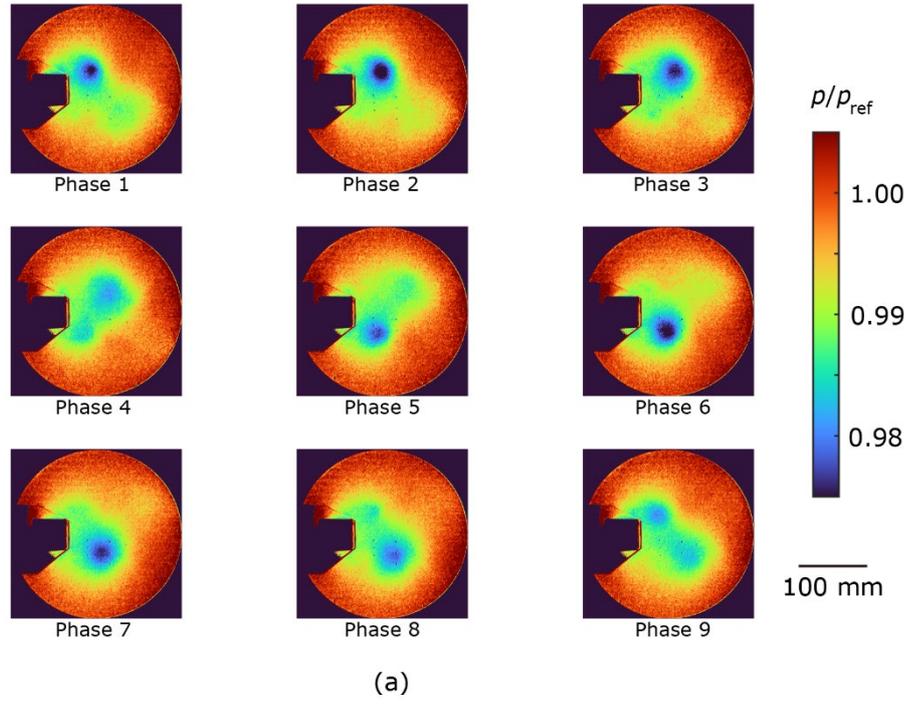

(a)

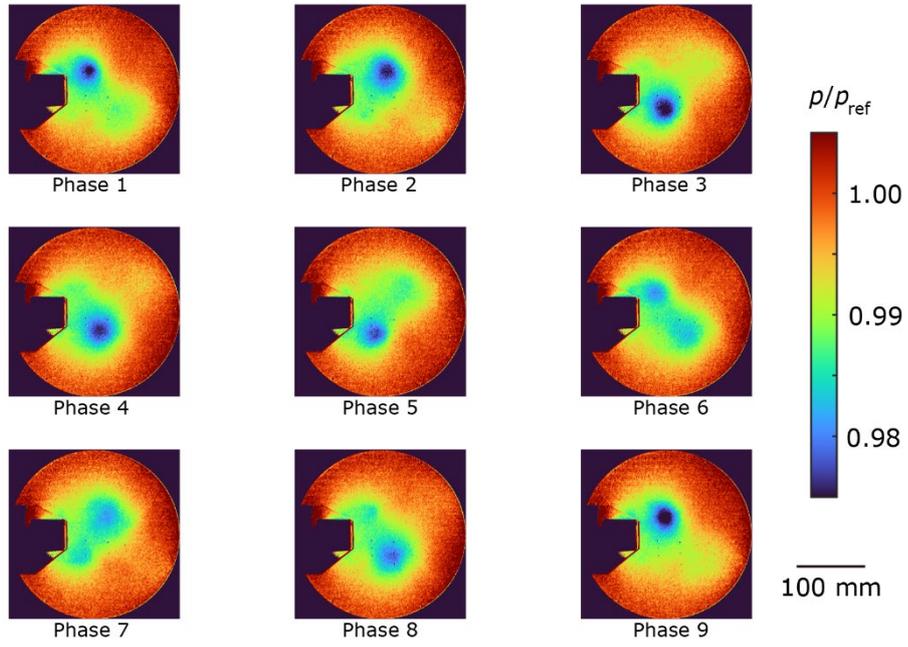

(b)

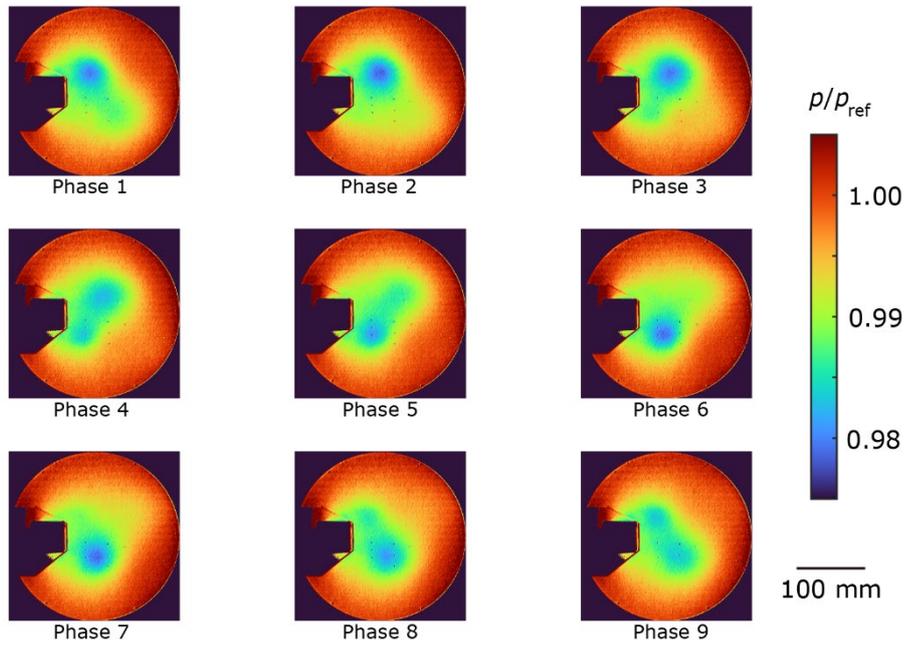

(c)

**Figure 4.** Ensemble averaged pressure distribution for (a) the proposed method, (b) the existing standard method (tslearn), and (c) the CIS method. Pressure $p$ is normalized by an atmospheric pressure $p_{\text{ref}}$.